\documentstyle[12pt]{article}

\begin{document}
\vspace*{.3cm}
\begin{flushright}
\large{CINVESTAV-FIS-13/96}
\end{flushright}
\begin{center}
\LARGE{\bf Non-leptonic $B$ decays involving tensor mesons}
 \end{center}
\vspace{.8cm}
\begin{center}
\Large G. L\'opez Castro$^1$ and J. H. Mu\~noz$^{1,2}$\\
\vspace*{.4cm}
{\normalsize{$^1$ \it  Departamento de F\'\i sica, Cinvestav del IPN, 
Apdo. \\ 
\vspace{-.2cm} \it  Postal 14-740, 07000 M\'exico, D.F., MEXICO. \\
$^2$ \it Departamento de F\'\i sica, Universidad del Tolima, \\
\vspace{-.4cm} \it A.A. 546, Ibagu\'e, COLOMBIA.}}
\vspace*{.4cm}
\end{center}

\thispagestyle{empty}
\centerline{ \bf Abstract}
\vspace{.3cm}
  Two-body non-leptonic decays of $B$  mesons into $PT$ and $VT$ modes 
are calculated using the non-relativistic quark model of Isgur {\em et 
al.}. The predictions obtained for $B \rightarrow \pi D^*_2,\ \rho D_2^*$ 
are a factor of $3\sim 5$ below present experimental upper limits. 
Interesting patterns are obtained for ratios of $B$ decays involving mesons 
with different spin excitations and their relevance 
for additional tests of forms factor models are briefly discussed.

 \vspace{.5cm} 

\newpage
\setcounter{page}{1}
\vspace{2cm}

  Weak non-leptonic decays of $B$ mesons involving mesons of intrinsic 
orbital momentum $l\geq 1$ in final states, are 
expected to be very suppressed \cite{pdg,gisw}. The experimental values for 
$B$ decay into orbitally excited charmed mesons, which are allowed at lowest 
order {\em 
via} external $W$-emission diagrams, exhibit the following suppressions 
\cite{pdg}:
 \begin{eqnarray*}
\frac{B(B^+ \rightarrow \overline{D^{*0}_2}\rho^+)}{B(B^+ \rightarrow 
\overline{D^0}\rho^+)} &\leq & 0.35 \\
\frac{B(B^+ \rightarrow \overline{D^{*0}_1}\rho^+)}{B(B^+ \rightarrow 
\overline{D^0}\rho^+)} &\leq & 0.104 \\
\frac{B(B^+ \rightarrow \overline{D^{*0}_1}\pi^+)}{B(B^+ \rightarrow 
\overline{D^0}\pi^+)} &= & 0.28 \\
\frac{B(B^0 \rightarrow D^{*-}_2\rho^+)}{B(B^0 \rightarrow 
D^-\rho^+)} &\leq & 0.63 . 
\end{eqnarray*}
Predictions for these $B$ decays are 
important because they would provide additional tests for the 
factorization hypothesis and form factor models used to describe 
exclusive modes in $B$ decays.
On the other hand, analogous ratios of $B$ decays involving lowest lying 
mesons ($l =0$), namely $B \rightarrow XV/XP$ ( $V$ and $P$ stand for vector 
and pseudoscalar mesons), such as 
\cite{pdg}: $B^+ \rightarrow \overline{D^0}\rho^+/\overline{D^0}\pi^+ = 
2.53,\ 
B^+ \rightarrow \overline{D^{*0}}\rho^+/\overline{D^{*0}}\pi^+ = 2.98,\
B^0 \rightarrow D^-\rho^+/D^-\pi^+ = 2.60$ and
$B^0 \rightarrow D^{*-}\rho^+/D^{*-}\pi^+ = 2.81$, show the expected  
behavior due to the three degrees of freedom of vector particles.

  Non-leptonic decays of $B$ and $D$ mesons involving 
scalar and tensor mesons have been calculated previously in a series of 
papers by Katoch and Verma \cite{kv1,kv2,kv3} using the non-relativistic 
quark model of Ref. \cite{gisw}. B decays into final states involving  
tensor mesons ($J^{PC}=2^{++}$) are not suppressed by phase-space 
considerations as in the 
case of $D$ decays. Actually, upper limits for $B$ decays into  $PT$ and 
$VT$ channels ($T$ denotes the tensor meson) 
at the level of $10^{-3} -10^{-4}$, both 
for Cabibbo-favored and Cabibbo-suppressed modes, are reported in the 
literature \cite{pdg}. Furthermore, according to the Particle Data Group 
(see 
p.99 in Ref. \cite{pdg}), the multiplet of tensor mesons is well established 
among the ones for the $q\overline{q}$ assignement for the case of four 
flavors.

   In this paper we compute the branching fractions for $B$ decays into $PT$ 
and 
$VT$ final states\footnote{ Note that the scalar+tensor final states are 
suppressed respect to $VT$ and $PT$ because the $<S|V^{\rm weak}_{\mu}|0>$ 
amplitude gives a further suppression factor.}
using the non-relativistic quark model of Isgur {\em et al.} \cite{gisw}. 
While there are not available 
predictions for the $VT$ decay modes, our results for the $PT$ modes 
differ quantitatively from those of Ref. \cite{kv1}. We present possible 
reasons for this discrepancy. 

   We start by introducing the effective weak hamiltonian for 
non-leptonic, Cabibbo favored, $b$ decays:
\begin{eqnarray}
{\cal H}_{eff} &=& \frac{G_F}{\sqrt{2}} \left\{ 
V_{cb}V_{ud}^*[a_1(\bar{c}b)(\bar{d}u)+ 
a_2(\bar{d}b)(\bar{c}u)] \right. \nonumber \\
&& \ \ \ \ \ \ \ \left. + 
V_{cb}V_{cs}^*[a_1(\bar{c}b)(\bar{s}c)+a_2(\bar{s}b)(\bar{c}c)] 
\right\}
\end{eqnarray}
where $(\bar{q}q')$ is a short notation for the V--A current, 
$G_F$ denotes the Fermi constant and $V_{ij}$ are the 
relevant Cabibbo-Kobayashi-Maskawa mixing factors. The numerical values 
$a_1=1.15$ and $a_2=0.26$, which are obtained from a fit to $B 
\rightarrow PP, VP$ decays \cite{kv1}, will be used throughout this paper.

   Since the internal $W$-emission diagrams are color suppressed, the 
leading contribution to the decays under consideration are given by the 
$W$-external diagram. Following Refs. \cite{kv1,kv2,kv3}, we write the decay 
amplitudes in the following general form:
\begin{equation}
{\cal M}(B \rightarrow PT) = i\frac{G_F}{\sqrt{2}} (CKM\ {\rm 
factors})(QCD\ {\rm factor}) f_P 
\varepsilon^*_{\mu\nu}p^{\mu}_Bp^{\nu}_B {\cal F}^{B 
\rightarrow T}(m_P^2) 
\end{equation} and 
\begin{equation}
{\cal M}(B \rightarrow VT) = \frac{G_F}{\sqrt{2}} (CKM\ {\rm 
factors})(QCD\ {\rm factor}) m_V^2f_V \varepsilon^{* \psi\tau}{\cal 
F}^{B\rightarrow T}_{\psi \tau}(m_V^2) 
\end{equation}
where 
\begin{eqnarray}
{\cal F}^{B\rightarrow T}(m_P^2)&=& k + b_+(m_B^2-m_T^2)+b_-m_P^2 \\
{\cal F}^{B\rightarrow T}_{\psi\tau}(m_V^2) &=& 
\varepsilon^*_{\mu}(p_B+p_T)_{\sigma}[ih\epsilon^{\mu\nu\sigma\rho}g_{\psi\nu}
(p_V)_{\tau}(p_V)_{\rho} +k\delta^{\mu}_{\psi}\delta^{\sigma}_{\tau} 
\nonumber \\
&& \ \ \ \ + b_+(p_V)_{\psi}(p_V)_{\tau}g^{\sigma\mu}].
 \end{eqnarray}
In the above expressions, $\varepsilon^*_{\mu}$ denotes the 
polarization four-vector of $V$, $\varepsilon^*_{\mu\nu}$ is the 
symmetric and traceless tensor describing the polarization of tensor 
mesons ($p_T^{\mu}\varepsilon_{\mu\nu}=\varepsilon_{\mu\nu}p_T^{\nu}=0$) 
and $p_i$ correspond to the four-momenta of the particles. The argument 
in the functions ${\cal F}^{B\rightarrow T}$ means that the form factors 
$h,\ k,\ b_+$ and $b_-$ should be evaluated at those values (namely, 
$m_P^2$ or $m_V^2$). The term $(QCD\ {\rm factor})$ in the amplitudes 
refers to either $a_1$ or $a_2$. Note 
that in the above expressions we have only one contribution to the decay 
amplitudes because the matrix element $<T|j_{\mu}^{weak}|0>$ vanishes  
identically.

   The hadronic matrix elements used in the previous expressions are defined 
as follows:
\begin{eqnarray}
<T|j^{\mu}|B> &=&ih\epsilon^{\mu\nu\phi\rho}\varepsilon^*_{\nu\alpha} 
p^{\alpha}_B(p_B+p_T)_{\phi}(p_B-p_T)_{\rho} 
+k\varepsilon^{*\mu\nu}(p_B)_{\nu} \nonumber \\ 
&& \ \ + \varepsilon^{*\alpha\beta}(p_B)_{\alpha}(p_B)_{\beta} 
[b_+(p_B+p_T)^{\mu}+b_-(p_B-p_T)^{\mu}] \\
 <P|A^{\mu}|0> &=& if_Pp_P^{\mu} \\ 
<V|V^{\mu}|0> &=& m_V^2f_V\varepsilon^{\mu}. 
\end{eqnarray}
The form factors $k,\ h,\ b_+$ and $b_-$ 
which describe the $B \rightarrow T$ transition, have been calculated in 
the non-relativistic quark model of Ref. \cite{gisw}. Note that $b_-$ 
gives a 
negligible contribution to $PT$ modes, while it does not appear at all in 
the decay amplitude for $VT$ modes because 
$p_V^{\mu}\varepsilon^{*}_{\mu}=0$. 

The central values for 
the decay constants of pseudoscalar mesons relevant for our 
calculations are (in GeV units): $f_{\pi}=0.131$, $f_{\eta_c}=0.384$ 
\cite{pdg,kv1} and $f_{D}=0.217,\ f_{D_s}=0.241$ (we assume the isospin 
symmetry 
relation $f_{D^0}=f_{D^+}$). The value for the $D_s$ decay constant 
\cite{ichep96} 
includes two recent determinations from the $D_s \rightarrow \mu \nu$ decay 
\cite{e653}, and $f_D$ is obtained using the theoretical prediction 
$f_D/f_{D_s}=0.90$ \cite{bls} based on a lattice calculation\footnote{ Other 
lattice 
computations are consistent with this result for $f_D/f_{D_s}$, while QCD 
sum rules predict a smaller value ($\approx 0.8$) for this ratio 
(see \cite{rb} for a summary of results).}. On the other hand, the 
dimensionless decay constants of vector mesons are:
$f_{\rho^-}= 0.2713$ (from $\tau \rightarrow 
\rho \nu_{\tau}$), $f_{J/\psi}=0.087$ (from $J/\psi \rightarrow 
e^+e^-$), while for the decay constants of $D^{*+}$ and 
$D^*_s$ we have estimated their values using the approximate scaling 
relation $m_V f_V 
\sim $ constant (see for example \cite{bsw}). Indeed, using $m_V f_V = 
0.231$ GeV ( which is obtained from a simple average of $f_{\rho^-},\ 
f_{K^{*-}},\ f_{J/\psi}$ and $f_{\Upsilon(1S}$), we obtain 
$f_{D^*}=0.1144$ and $f_{D^*_s} = 0.109$.

  The expressions for the decay amplitudes in the exclusive modes of $B$ 
decays are given explicitly in Table 1  for each of the 
$VT$ modes allowed in the leading approximation; our expressions for the 
$PT$ amplitudes coincide with those given in Table 1 of Ref. \cite{kv1}.

   From the previous expressions we obtain the following decay rates for 
the $PT$ modes: 
\begin{equation}
\Gamma (B \rightarrow PT) = |{\cal A}(B\rightarrow PT)|^2 \left( 
\frac{m_B}{m_T}\right)^2 \frac{|\vec{p}_P|^5}{12\pi m_T^2}
\end{equation}
where
\begin{equation}
{\cal A}(B\rightarrow PT) =\frac{G_F}{\sqrt{2}} (CKM\ {\rm factors})(QCD\ 
{\rm factor}) f_P {\cal F}^{B\rightarrow T}(m_P^2), \nonumber
\end{equation}
and the expression for the decay into the $VT$ channels is given by:
\begin{eqnarray}
 \Gamma (B \rightarrow VT) &=& \frac{G_F^2}{48\pi m_T^4}(CKM\ {\rm 
factors})^2 (QCD\ {\rm factor})^2 \nonumber \\
&& \ \ \ \ \cdot \ m_V^2f_V^2 [ \alpha 
|\vec{p}_V|^7+ \beta |\vec{p}_V|^5 + \gamma |\vec{p}_V|^3 ]
 \end{eqnarray}
where $\alpha,\ \beta$ and $\gamma$ are quadratic functions of the form 
factors evaluated at $q^2=m_V^2$, namely:
\begin{eqnarray*}
\alpha &=& 8 m_B^4b_+^2 \\
\beta &=& 2m_B^2 [6m_V^2m_T^2h^2+2(m_B^2-m_T^2-m_V^2)kb_++k^2] \\
\gamma &=& 5m_T^2m_V^2 k^2 .
\end{eqnarray*}
In the above expressions for the decay rates, $\vec{p}_{V(P)}$ denotes 
the three-momentum of the $V(P)$ meson in the $B$ rest frame 
($\vec{p}_{V(P)}$ depends, of course, on the specific decay considered).

  In order to obtain the unpolarized rates given in Eqs. (9,11), we have 
used the following expression for the sum over polarizations of the 
tensor meson:
\begin{eqnarray}
P_{\mu\nu\alpha\beta} &=& \sum_{\lambda} \varepsilon_{\mu\nu}(p,\lambda) 
\varepsilon^*_{\alpha\beta}(p,\lambda) \nonumber \\
&=& \frac{1}{2}(\theta_{\mu\alpha}\theta_{\nu\beta}+ 
\theta_{\mu\beta}\theta_{\nu\alpha}) -\frac{1}{3} 
\theta_{\mu\nu}\theta_{\alpha\beta},
 \end{eqnarray}
where $\theta_{\mu\nu} \equiv -g_{\mu\nu} + p_{\mu}p_{\nu}/m_T^2$. 
Observe that $P_{\mu\nu\alpha\beta}$ satisfies the following identities:
 \begin{eqnarray*}
&& P_{\mu\ \ \alpha\beta}^{\ \mu}=P_{\mu\nu\alpha}^{\ \ \ \alpha}=0 \\
&& P_{\mu\nu\alpha\beta} \epsilon^{\alpha\beta} = \epsilon_{\mu\nu} \\
&& P_{\mu\nu\rho\sigma}P^{\rho\sigma}_{\ \ \alpha\beta} = 
P_{\mu\nu\alpha\beta}. 
\end{eqnarray*}

   Now, let us address some comments on the expressions for the decay 
rates given in Eqs. (9,11):\\
(1) The decay rate in the $PT$ mode, Eq. (9), contains an 
additional factor $(m_B/m_T)^2$ with respect to Eq. (6) in Ref. \cite{kv1}. 
This would enhance the prediction for the $PT$ modes by a factor 
of around 4 $\sim$ 17, depending on the specific decay mode, with respect to 
the predictions of Ref. \cite{kv1}. The 
origin for this discrepancy could be the expression used for the sum over 
polarizations of tensor mesons. Note that using Eq. (12), we reproduce the 
results given in Refs. \cite{tensor} for the strong decay rates of tensor 
mesons. \\
(2) The $|\vec{p}|^5$-dependence in Eq. (9) indicates that only the $l=2$ 
wave is allowed for the $PT$ system, while in the $VT$ decay modes the 
$l=1,\ 2$ and $3$ are simultaneously allowed, as expected.

  The numerical values for the corresponding branching ratios are given 
in Table 2, where we have used the $B$ lifetimes values reported in Ref. 
\cite{pdg}, and $|V_{bc}|= 0.041 ,\ |V_{cs}| = 0.975$ and $|V_{ud}| =0.9736$ 
\cite{pdg}. When we computed the rates involving an isoscalar tensor meson 
($f_2$ or $f'_2$), we have used an octet-singlet mixing angle 
$\theta_T =28^o$ or equivalently the deviation from ideal mixing $\phi_T 
\equiv  \arctan(1/\sqrt{2})-\theta_T \approx 8.3^0$, as in Ref. \cite{kv1}.

   From Table 2, we can raise  the following conclusions:\\ 
(1) The values for the $PT$ branching fractions are larger by a factor of 
2.6 $\sim$ 15 than those given in Ref. \cite{kv1}. The reason for this 
enhancement has been explained above; the smallest enhancement factor 
corresponds to the $B \rightarrow D_s D_2^*$ decay mode, while the 
largest value corresponds to $B \rightarrow \eta_c K^*_2$. Note that part of 
the difference between our 
results and those in Ref. \cite{kv1} arises from the smaller values that 
we use for the $D,\ D_s$ decay constants.\\
(2) When we consider the ratio for the decay modes $VT/PT$, $V$ and 
$P$ having identical quark content, we obtain
\begin{equation}
\frac{\Gamma(B\rightarrow VT)}{\Gamma(B\rightarrow PT)} = 
\frac{m_V^2f_V^2 \{ \alpha |\vec{p}_V|^7 +\beta|\vec{p}_V|^5 
+\gamma |\vec{p}_V|^3 \} }{2[m_Bf_P {\cal F}^{B \rightarrow 
T}(m_P^2)]^2 |\vec{p}_P|^5}. 
\end{equation} 
This ratio
turns out to be independent of $a_{1, 2},\ G_F$ and Kobayashi-Maskawa 
mixing factors and would provide a clean test of the factorization 
hypothesis and of the ISGW model \cite{gisw}. In the last column of Table 
2 we 
show the ratios corresponding to Eq. (13). The results indicate that the 
ratios $VT/PT \sim 3$ for processes which amplitudes are 
proportional to $a_1$, which are normally expected due 
to the three degrees of freedom of vector particles. Note, however, that 
for the other processes, $(V T)/(PT)$ differs substantially from 3 because 
these decays are not allowed at tree level through the emission of 
external $W$ (they proceed {\em v\'\i a} amplitudes proportional to 
$a_2$).\\ 
 (3) Our predictions for the decays $B^- \rightarrow 
(\pi^-,\ \rho^-)D_2^{*0}$ and $\overline{B^0} \rightarrow (\pi^-,\ 
\rho^-)D_2^{*+}$ are below the present experimental upper limits reported 
in \cite{pdg} by a factor of 3$\sim$5. This offers optimistic prospects for 
their measurements and for additional tests of the model of Ref. 
\cite{gisw}.\\
(4) If we use the experimental measurements of $B \rightarrow J/\psi 
K^*(892)$ and $B \rightarrow D_s D^*$ \cite{pdg}, we can calculate the 
following 
ratios from Table 2: $B^- \rightarrow (J/\psi K_2^{*-})/(J/\psi 
K^{*-}(892)) \approx \overline{B^0} \rightarrow (J/\psi K_2^{*0})/(J/\psi 
K^{*0}(892)) 
\approx B^- \rightarrow (D_s^-D_2^{*0})/(D_s^-D^{*0}) \approx B^0 
\rightarrow 
(D_s^+D_2^{*-})/(D_s^+D^{*-}) \approx 0.05$. Thus, the 
ratios $(B\rightarrow XT)/(B\rightarrow XV)$, which measure the effects 
of the dynamics due the orbital excitation $ V\rightarrow T$, turns out 
to be very suppressed in the $\Delta s=-1$ channels.

  In conclusion, in this paper we have computed the Cabibbo-favored 
decays of $B$ mesons into $VT$ and $PT$ final states, using the 
non-relativistic quark model of Isgur {\em et al.} \cite{gisw}. Our results 
for 
the $\pi D_2^*$ and $\rho D_2^*$ decay channels are a factor of 3$\sim$5 
below present experimental upper limits \cite{pdg}. Our results exhibit 
interesting 
patterns for final states mesons of higher orbital momentum excitations,
 which would offer additional tests for the factorization hypothesis and 
the form factor models for exclusive $B$ decays.

\

{\bf Acknowledgements}   

  The authors would like to acknowledge financial support from Conacyt 
(GLC) and Colciencias (JHM).

\newpage

\

\begin{center}
\begin{tabular}{|c|c|c|}
\hline
& Process & Amplitude $\times (V_{cb}V_{ud}^*)$ \\
\hline\hline
 & $B^- \rightarrow \rho^-D_2^{*0}$ & $a_1f_{\rho^-}m_{\rho^-}^2{\cal F}^{B 
\rightarrow D_2^*}_{\mu\nu} (m_{\rho}^2)$ \\
 & $B^- \rightarrow D^{*0}a_2^-$ & $a_2f_{D^{*0}}m_{D^{*0}}^2{\cal F}^{B 
\rightarrow a_2}_{\mu\nu} (m_{D^*}^2)$ \\
$\Delta s=0$ & $\overline{B^0} \rightarrow \rho^-D_2^{*+}$ & 
$a_1f_{\rho^-}m_{\rho^-}^2{\cal F}^{B \rightarrow D_2^*}_{\mu\nu} 
(m_{\rho}^2)$ \\
 & $\overline{B^0} \rightarrow D^{*0}a_2^0$ & $-\ 
a_2f_{D^{*0}}m_{D^*}^2{\cal 
F}^{B \rightarrow a_2}_{\mu\nu} (m_{D^*}^2)/\sqrt{2}$ \\
 & $\overline{B^0} \rightarrow D^{*0}f_2$ & $ a_2f_{D^{*0}}m^2_{D^*}\cos 
\phi_T {\cal F}^{B \rightarrow f_2}_{\mu\nu} (m_{D^*}^2)/\sqrt{2}$ \\
 & $\overline{B^0} \rightarrow D^{*0}f'_2$ & $ a_2f_{D^{*0}}m^2_{D^*}\sin 
\phi_T {\cal F}^{B \rightarrow f'_2}_{\mu\nu} (m_{D^*}^2)/\sqrt{2}$ \\
\hline\hline
& Process & Amplitude $\times (V_{cb}V_{cs}^*)$ \\
\hline\hline
& $B^- \rightarrow D_s^{*-}D_2^{*0}$ & $a_1 f_{D_s^*}m^2_{D^*_s}{\cal 
F}^{B\rightarrow D_2^{*0}}_{\mu\nu}(m^2_{D_s^*})$ \\
$\Delta s=-1$& $B^- \rightarrow J/\psi K_2^{*-}$ & $a_2 
f_{J/\psi}m^2_{J/\psi} {\cal F}^{B\rightarrow K_2^*}_{\mu\nu}(m^2_{J/\psi})$ \\
& $\overline{B^0} \rightarrow D_s^{*-}D_2^{*+}$ & $a_1 f_{D_s^*}m^2_{D_s^*} 
{\cal F}^{B\rightarrow D_2^*}_{\mu\nu}(m^2_{D_s^*})$ \\
& $\overline{B^0} \rightarrow J/\psi\overline{K_2^{*0}}$ & $a_2 
f_{J/\psi}m^2_{J/\psi} 
{\cal F}^{B\rightarrow K_2^*}_{\mu\nu}(m^2_{J/\psi})$ \\
 \hline
\end{tabular}
\end{center}
\begin{center}
Table 1. Decay amplitudes for the CKM-favored $B \rightarrow VT$ channels 
with $\Delta s=0,\ -1$. The tabulated amplitudes must be multiplied 
by $(G_F/\sqrt{2})\varepsilon^{*\mu\nu}$.
 \end{center}

\newpage
\

\begin{center}
\begin{tabular}{|c|c|c||c|c||c|}
 \hline
&$B \rightarrow PT$ & BR($B \rightarrow PT$) & $B\rightarrow VT$ & BR($B 
\rightarrow VT$) & $VT/PT$ \\ 
\hline\hline
& $B^- \rightarrow \pi^- D^{*0}_2$ & $4.07 \cdot 10^{-4}$ & $B^- 
\rightarrow \rho^- D^{*0}_2$ & $1.13 \cdot 10^{-3}$ & $ 2.79$ \\
& $B^- \rightarrow D^0 a^{-}_2$ & $1.35 \cdot 10^{-5}$ & $B^- 
\rightarrow D^{*0} a^{-}_2$ & $ 2.23 \cdot 10^{-5}$ & $1.65$\\
& $\overline{B^0} \rightarrow \pi^- D^{*+}_2$ & $4.06 \cdot 10^{-4}$ & 
$\overline{B^0} 
\rightarrow \rho^- D^{*+}_2$ & $1.13 \cdot 10^{-3}$ & $2.80$\\
$\Delta s\!=\! 0 $ & $\overline{B^0} \rightarrow D^0 a^{0}_2$ & $6.79 \cdot 
10^{-6}$ & 
$\overline{B^0} \rightarrow D^{*0} a^{0}_2$ & $ 1.13 \cdot 10^{-5}$ & 
$1.66$\\ & $\overline{B^0} \rightarrow D^0 f_2$ & $7.34 \cdot 10^{-6}$ & 
$\overline{B^0} \rightarrow D^{*0} f_2$ & $1.17 \cdot 10^{-5}$ & $1.60$\\
& $\overline{B^0} \rightarrow D^0 f'_2$ & $8.73 \cdot 10^{-8}$ & 
$\overline{B^0} \rightarrow D^{*0} f'_2$ & $2.01 \cdot 10^{-7}$ & 
$2.30$\\ 
\hline
& $B^- \rightarrow D_s^- D^{*0}_2$ & $2.69 \cdot 10^{-4}$ & $ 
B^-\rightarrow D^{*-}_sD^{*0}_2$ & $1.05 \cdot 10^{-3}$ & $3.91$\\
$\Delta s\! =\! -1$ & $B^-\rightarrow \eta_c K^{*-}_2$ & $ 8.08 \cdot 
10^{-6}$ & $B^- \rightarrow J/\psi K^{*-}_2$ & $ 7.62 \cdot 10^{-5}$ & $ 
9.43$ \\
& $\overline{B^0} \rightarrow D_s^- D^{*+}_2$ & $2.69 \cdot 10^{-4}$ & $ 
\overline{B^0}\rightarrow D^{*-}_sD^{*+}_2$ & $1.05 \cdot 10^{-3}$ 
&$3.91$\\
& $\overline{B^0}\rightarrow \eta_c \overline{K^{*0}_2}$ & $ 7.53 \cdot 
10^{-6}$ & $\overline{B^0} \rightarrow J/\psi \overline{K^{*0}_2}$ & $ 7.55 
\cdot 10^{-5}$ & $10.03$ \\
\hline
\end{tabular}
\end{center}
\begin{center}
Table 2. Branching ratios for Cabibbo-favored $B\rightarrow PT,\ VT$ 
decays with $\Delta s=0,\ -1$. Last column shows the ratio for $VT/PT$ 
branching ratios.
 \end{center}
\

\end{document}